
%
%
\def\draftversion{N}   
\def\preprint{Y}       

\if \draftversion Y


\fi

\def\today{\ifcase\month\or
	January\or February\or March\or May\or April\or June\or
	July\or August\or September\or October\or November\or December\fi
	\space\number\day, \number\year}
\input jnl \refstylenp \def\cit#1{[#1]}

\if \preprint Y  \twelvepoint\oneandathirdspace \fi

\if \draftversion Y \rightline{\today{}} \fi

\def\sla{\raise.15ex\hbox{$/$}\kern-.57em}

\def\acta{\journal Acta Physica Polonica,}

\def\lag{{\cal L}}

\if \draftversion N \input reforder \input eqnorder \citeall\cit \fi

\if \preprint Y \rightline{RU-92-58} \fi

\title Infinitely many regulator fields for chiral fermions
\vskip 1cm
\author Rajamani Narayanan and Herbert Neuberger
\affil
\vskip 1.cm
\centerline{\it Department of Physics and Astronomy}
\centerline{\it Rutgers University}
\centerline{\it Piscataway, NJ 08855--0849}
\vskip 1.5 cm

\goodbreak

\abstract  We show that two recent independent proposals for regularizing a
chiral
gauge theory stem from one common trick. If the anomaly free complex
representation
carried by the right handed fermi--fields is $r$ one constructs a vector like
theory with flavored
right handed fermionic matter in $r+\bar r$ but with a mass matrix of the
order of the cutoff and having an index equal to unity in an infinite
dimensional
flavor space. We present a Pauli--Villars realization of the trick that is
likely to work to all
orders in perturbation theory and a lattice version which is
argued to produce the correct continuum leading order
fermionic contribution to the vacuum polarization tensor and  readied for
further perturbative checks.
\endtitlepage

\if \preprint N \doublespace \fi

\head{Introduction}

This paper  addresses the problem of regularizing a nonabelian chiral gauge
theory with
right handed fermionic matter in an anomaly free complex representation of the
gauge group.
For definiteness we work in four dimensions and assume the model to be
asymptotically
free so that according to standard lore a continuum interacting theory with no
free parameters
exists.
Following \cit{ROME} we refer to this continuum theory
as the target theory; a concrete
example for a target theory is $SO(10)$ with righthanded fermions in the
spinorial representation.
We shall focus on a particular trick that opens up the possibility for a gauge
invariant regularization
at the expense of introducing an infinite set of fermionic regulator fields.
Two recent
independent proposals \cit{FSLAV, KAPLAN}, although quite different in
appearance, can be built around this trick.

We start in Minkowski space. The gauge group index will be suppressed and plays
no role in most
of what follows. To each righthanded fermion of the target theory we attach a
tower consisting of
righthanded and lefthanded fermionic fields labeled by a new index that lives
in a new
internal space. The total fermion number of each tower is conserved.

We wish to give Dirac masses of the order of the cutoff to all the extra fields
and,
at the same time, make the theory look vector like and gauge invariant. We
write the Lagrangian
density as:
$$
\eqalign{\lag=&\lag_1 +\lag_g (A),~~~\lag_1 =i\bar\psi (\sla\partial -ig\sla A
)\psi +\bar\psi (MP_R +M^{\dagger} P_L )\psi\cr
P_R =&{{1+\gamma_5 }\over{2}},~~~P_L ={{1-\gamma_5 }\over{2}},~~~A_{\mu}
=A^a_{\mu} T^a,~~~tr(T^a T^b )=\delta^{ab}\cr}
 \eqno(LAG)$$
To make the theory look vector like we need to identify the internal spaces
associated with
the lefthanded and righthanded fields and $M$ becomes an operator acting in
this space. The free
fermionic theory will have massless righthanded particles equal in number to
$n_R
=dim(Ker(M))$ and massless lefthanded particles equal in number to $n_L
=dim(Ker(M^{\dagger}))$.
If the internal space has a finite dimension, $n_R = n_L$, and we cannot get to
the target theory.
To make $n_R =1$ and $n_L =0$ we need an infinite dimensional space; a simple
example is
$M=\Lambda {\bf a}$ where $\Lambda$ is the cutoff and ${\bf a}$ is a bosonic
annihilation operator.

If $M$ were a finite matrix  and $M'$ its restriction to the subspace
orthogonal to $Ker(M)$ the parity breaking appearance of  eq.~\(LAG) could be
eliminated by
$\psi\rightarrow\exp(i\alpha\gamma_5 )\psi$ where $\alpha$ is a hermitian
matrix satisfying
$\exp(2i\alpha )=(M'M'^{\dagger})^{-1/2}M'$. This doesn't work in our case and
parity
is not conserved. However, one can define $U=M^{\dagger} (MM^{\dagger})^{-1/2}$
and if one
``forgets'' that $U$ is only formally unitary\footnote{*}{$U^{\dagger}U=1$ but
$UU^{\dagger}=M^{\dagger}(MM^{\dagger})^{-1}M=1-Q$
where $Q$ is the projector on the zero eigenspace of $M$.}  the theory
looks vector like and easy to regularize.

One can easily find more examples of $M$ but one should try to keep the
structure as simple as possible.
Before we had $M/\Lambda =\exp(-{{s^2}\over {2}} )\partial_s \exp({{s^2}\over
{2}} )$ where $s\in {\rm R}$ is the internal space ``coordinate'' of the
harmonic oscillator. This can be generalized to
$M/\Lambda =\exp(-F(s) )\partial_s \exp(F(s) )$ where $\exp(-F(s))$ is square
integrable and the
internal space is the space of square integrable functions on the line. This
choice assures that
the kernels of $M$ and its adjoint will have the correct dimensions. One can
also generalize to an
internal space corresponding to a complex coordinate with $M/\Lambda
=\exp(-F(z,\bar z) )\partial_z \exp(F(z,\bar z) )$. In the last example $M$
represents one chirality component of a Dirac particle living in two Euclidean
dimensions in an external magnetic field and the flux of that field must be
appropriately adjusted to get the correct number of zero modes. The other
chirality component is represented by $M^{\dagger}$.

In all the examples above the internal space corresponds to additional internal
dimensions and
ultraviolet behavior might be worse than in the bare theory because there are
more single particle
states at high energies. Therefore, regularization has to deal with all the
original divergences
and also potential new ones. If the spectrum of $MM^{\dagger}$ is discrete
the fields can be labeled by $0,1,2,...$ and it is natural to view the extra
fields as Pauli-Villars
regulator fields whose masses and statistics are adjusted so that everything is
ultraviolet finite
to any order in perturbation theory. If the spectrum of $MM^{\dagger}$ has an
unbounded
continuous component it is more natural to first try to get rid of the extra
ultraviolet divergences
induced by the internal space and this is easiest done by latticizing the
variable $s$ and thereby
bounding the spectrum of $MM^{\dagger}$ from above.  It is natural then to also
latticize real space
and try to get a non--perturbative definition of the target theory with hopes
of addressing such
long standing vexing basic questions about chiral gauge theories as, for
example, what their
spectrum is. We explore both possibilities in this paper.

We first analyze the Pauli--Villars case and show that it
essentially corresponds to the suggestion of \cit{FSLAV}. Our point of view
seems quite different
from the one adopted there, to the extent we understand it; our final formulae
can be made to differ
only in some of the details and we think that the differences are
insignificant. The indications are
that the method works for an anomaly free situation but breaks down when the
gauge group
is represented by fermions in an anomalous representation. Ungauged, global
nonabelian
symmetries can be preserved only if they are non--anomalous.
Next we discuss the lattice case
and show that it is closely related to the proposal of \cit{KAPLAN}. Our
version will differ from this
proposal in the manner in which the fermions are coupled to the gauge fields;
we shall
not introduce extra gauge or other bosonic fields beyond those present  in the
target theory. Our
study will be somewhat preliminary: we shall provide a necessary tool that was
missing until now
which is an explicit formula for the fermionic propagator and shall argue that
the vacuum polarization induced by fermions has the right
continuum limit. We hope to come back in the future
with a more detailed perturbative study of the lattice version; as far as we
can see now the
outlook for this approach is better than anything we have seen in years, but a
completely well
defined construction is still lacking and more work is needed.

\head{Pauli--Villars}

We assume that $MM^{\dagger}$ and $M^{\dagger}M$ have discrete nondegenerate
spectra. In addition
we assume the existence of a ``parity'' operator $\hat S$ which anticommutes
with $M,$ is hermitian and squares
to unity. The pure gauge part is regularized in some way, unspecified except
that it preserves ordinary, continuous, space--time structure and gauge
invariance.
For example, a higher derivative regularization, as defined in \cit{FADSLAV},
does this. We are
considering in this subsection only
perturbation theory and any loop with at least a gauge or ghost internal line
is assumed to be made finite
by the extra regularization. Thus, all we need to do is to
construct a gauge invariant regularization of the fermionic
determinant in an arbitrary external gauge field. Not surprisingly, the overall
regularization scheme will only work if the
representation of the gauge group carried by the righthanded fermions is
anomaly free.

The free fermion propagator is:
$$
[\sla p -MP_R -M^{\dagger}P_L]^{-1}=(\sla p +M^{\dagger}) P_R {{1}\over{p^2 -
MM^{\dagger}}}  +
(\sla p +M)P_L{{1}\over{p^2 - M^{\dagger}M}} \eqno(FREEPROP)$$
We choose the statistics to be well defined for fields that are eigenstates of
$\hat S \gamma_5$. In this way
the Lagrangian only couples fields of the same statistics: the $\sla\partial
-\sla A$ term is unity in internal space and
couples only fields of the same handedness while the mass terms only couple
fields of opposite handedness and
opposite $\hat S$ parity. A fermion loop is given by tracing over internal
space, spinor index, gauge group index and
integrating over the loop momentum. Statistics is taken into account by
inserting at some point one statistics factor equal to $-S$. To
get the right target theory one needs to pick the $\hat S$ parity of the zero
eigenvector of $M$ as unity.

As an example consider the lowest order contribution to the vacuum
polarization, given by $<j_{\mu ,a}j_{\nu ,b}>=\delta_{ab }\Pi_{\mu\nu}$. After
a certain amount of algebra one obtains:
$$
\Pi_{\mu\nu} (p) =\int {{d^4 k}\over{(2\pi)^4}}  \bigg [  Tr \big [ \hat S
(1-{{Q}\over{2}} )  {{1}\over {\sla q_+  -\sqrt {M^{\dagger}M}}}\gamma_{\mu}
{{1}\over {\sla q_-  -\sqrt {M^{\dagger}M}}}\gamma_{\nu} \big ] +{1\over{2}}tr
[ {{1}\over {\sla q_+}}\gamma_{\mu} {{1}\over {\sla q_- }}\gamma_{\nu} \gamma_5
 ] \bigg ]
\eqno(VACPOL)$$
$q_{\pm} = k\pm {{p}\over{2}} $ and $Tr$ denotes a trace over everything while
$tr$ sums over spinor indices.

The $\gamma$ matrix trace in the last term is proportional to
$\epsilon_{\mu\nu\alpha\beta} q_+^\alpha
q_-^\beta $; under  $k\rightarrow -k$ the term is odd and hence the momentum
integral over it vanishes.
This formal argument\footnote{*}{The last term in eq.~\(VACPOL) does not
converge by power counting.} can be easily made rigorous.
Once the first term in eq.~\(VACPOL) is arranged to be finite the
transversality of the vacuum polarization tensor will
be assured by the usual (no longer formal) argument.

To understand finiteness let us consider a slightly
more general type of loop integral $\int {{d^4 k}\over{(2\pi )^4}} I(k)$
with
$$
I(k) = polynomial(q_j ) Tr \bigg [ \hat S \prod_{j=1}^n \bigg ( {{1}\over{q_j^2
-M^{\dagger}M}}\bigg )(1-
{{Q}\over{2}})\bigg ]
\eqno(GENINT)$$
where $q_j =q_{j-1} + p_j ,~~j=2,3,....n~~q_1 =k$ and the $p_j$ are external
momenta that sum to zero. The trace
is over the internal space. Pick the basis of the space as $|0> , |E>$ with
$M^{\dagger}M |0>=0$ and $M^{\dagger}M|E> = E |E>, E>0$. We assume that there
are no degeneracies. Since $\hat S$ commutes with
$M^{\dagger}M$ we have $\hat S|E>=\eta_E |E>, \hat S|0> =|0> $ and $\eta_E =\pm
1$. After exponentiating each
of the denominators with the help of a parameter $t_j$ and rotating to
Euclidean space it is clear that the question
of convergence is settled by the behavior of the trace when all $t_j$'s are
scaled to zero. Let $\sum t_j =\tau$; we
have to consider then the behavior
of
$$
Sum= 2\sum_E \eta_E \exp (-E\tau) +1\eqno(SUM)$$
Because of gauge invariance and absence of anomalies
we only need to tame the logarithmic divergence
of a few diagrams. It is then sufficient to make
the expression in eq.~\(SUM) vanish linearly when $\tau$ goes to zero. This is
easily achieved with $M=\Lambda {\bf a}$
and $\hat
S = (-1)^{{\bf a}^{\dagger}{\bf a}}$ leading to $Sum=\tanh (\Lambda^2 \tau )$.
Better control in the ultraviolet is achievable; for example one may be
interested in the determinant of $\sla\partial -\sla A$ in higher even
dimensions. An arbitrary
degree of divergence can be tamed
by choosing $M=\Lambda \sqrt {{\bf a}
{\bf a}^{\dagger}}{\bf a}$ and the same statistics operator as above.
Now the sum becomes $\theta_4 (0|e^{-\Lambda^2 \tau})$ and the Jacobi identity
ensures that it vanishes
faster than any power when $\tau$ approaches zero. This latter choice has been
made in \cit{FSLAV}.

At any order we shall have to deal with one extra piece, similar to the one we
discarded in the calculation of the
vacuum polarization, before we can restrict our attention to sums of the kind
shown in eq.~\(GENINT). At higher orders these
extra terms converge by power counting but diagrams with
three and four vertices may be divergent
and untamable. Our investigations indicate that
if the theory is anomaly free no problems are encountered and everything goes
through as for the vacuum polarization graph.

Essentially, the Pauli-Villars regularization removes all divergences in the
parity conserving
part of the fermionic determinant but does not touch the parity odd part of the
logarithm of this determinant. The latter
must be finite and {\it unambiguous} ``by itself'' without any help
from the Pauli--Villars regulator fields and this happens only in the anomaly
free case.

It is important to make sure that the counter terms
will not affect the zero mode structure of the mass
matrices; otherwise the target theory would not be attained.
This is easily checked to leading order and turns just into a matrix
generalization of the well known
$\delta m_f \propto m_f$ rule. Here this works is an obvious way because the
internal space is totally
decoupled from the real space and the spinor and group indices. However,
the mechanism that protects the desired zero mode structure from radiative
corrections is a
deeper one and this becomes an essential feature when we turn to the lattice
where such a decoupling cannot be achieved because of the well known ``lattice
fermion doubling''.
The lattice version
therefore also avoids infinite amount of fermion mass
fine tuning \cit{SVETLONG}.

\head{Lattice}
A simple choice of the mass matrix in the lattice case is based on the ``wall''
realization of $M$. Its
continuum form is $M/\Lambda =\partial_s +f(s),~~f(s)=F'(s)$ with $f$
increasing monotonically from a
negative asymptotic value to a positive one. In particular, this means that the
internal manifold is not
compact. $M$ will have a zero eigenfunction and $M^{\dagger}$ will have none;
this feature
is stable under deformations that do not change the large
$|s|$ behavior \cit{WITTEN}: $MM^{\dagger}$ and $M^{\dagger}M$ have identical
spectra for
nonvanishing eigenvalues and therefore any small deformation of $M$ cannot move
the zero mode
of $M^{\dagger}M$ up without pairing it up with an eigenstate of
$MM^{\dagger}$ of exactly the same energy; however we assume that there is a
finite gap
between the zero mode and the rest of the spectrum and a small deformation
cannot drag down
an eigenstate of $MM^{\dagger}$ far enough.
It is hoped that radiative
corrections will be well behaved in this sense if the theory is anomaly free.
Plugging the
``wall'' form of $M$ into
eq.~\(LAG) leads to the interpretation of  $s$ as a fifth dimension, the
derivative part
of $M$ and $M^{\dagger}$ (i.e. ${{M-M^{\dagger}}\over 2}$) combining with the
kinetic energy into a five dimensional Dirac operator,
and the $f(s)$ dependent part (i. e. ${{M+M^{\dagger}}\over 2}$) becoming a
variable mass term.

The spectrum of $MM^{\dagger}$
is unbounded and to cure this we replace inner
space by an infinite lattice in both directions. $s$ now denotes
the discrete integer labeling the sites
on this internal lattice and
$M$ is replaced by a first order difference operator:
$M_{s,s'}=\delta_{s+1,s'} -a(s)\delta_{s,s'}$. If $a(s)$ approaches $a_+$ with
$|a_+|<1$ when $s\rightarrow\infty$ and $a_-$ with
$|a_-|>1$ when $s\rightarrow -\infty$ the spectral properties we need are
preserved. For definiteness we choose $a^0 (s) = 1-m_0{\rm sign} (s+{1\over
2})$ with $0 <m_0 < 1$ and denote the corresponding $M$ by $M^0$.

Putting an Euclidean version (with hermitian $\gamma$ matrices) of eq.~\(LAG)
on the lattice we
obtain:
$$
\eqalign{
-\lag_E =&- \beta \sum_{x,\mu} \Re\{ tr [U_\mu (x) U_\nu (x+\mu )
U_{\mu}^{\dagger}(x+\nu ) U_{\nu}^{\dagger} (x) ] \}+ \sum_{x,\mu} {1\over {2}}
\bar\psi (x) \gamma_\mu
[ U_\mu (x) \psi (x+\mu) \cr &-U_\mu^{\dagger} (x-\mu )\psi (x-\mu )] + \sum_x
\bar\psi (x) [P_R \big ( M(\nabla_U )\psi \big ) (x)+ P_L \big ( M^{\dagger}
(\nabla_U ) \psi \big )(x) ]\cr
& \big ( \nabla_U  \psi \big ) (x) =\sum_\mu [U_\mu (x) \psi(x+\mu )
+U_\mu^{\dagger} (x-\mu )\psi (x-\mu ) -2\psi (x) ]\cr}
\eqno(EUCACT)$$
For noninteracting fermion fields in Fourier space, $M$ depends parametrically
on the lattice momentum
$p$ ($-\pi < p_\mu \le \pi$ ) with $\nabla_{U=1}\equiv\nabla =-4\sum_{\mu}
\sin^2 {{p_\mu }\over {2}}$ so that all the unwanted potential zeros of the
propagator are moved up in energy by Wilson
mass terms.  We have to require that
for $\nabla$ in a region close to $p_\mu =0$ $dim(Ker(M(\nabla )))=1$ but for
$\nabla \le -4$ $dim(Ker(M(\nabla )))=0$. We also require that $M^\dagger$ have
no zero modes in regions around the momenta $p$ with $\sin (p_\mu ) =0$ for any
$\mu$. We need to change the
asymptotic behavior in $s$ so that  as $\nabla$ varies over the
Brillouin zone the spectra of $M(\nabla )$ and $M^{\dagger}(\nabla )$ change as
follows:
for small momenta that satisfy $-\nabla < h$ (an example for $h$ will be
provided later),
$M$ has a normalizable zero mode
and the spectrum of $M(\nabla )M^{\dagger}(\nabla )$ is continuous with a
finite gap
separating it from zero. On the boundary of this small momentum region,
$-\nabla = h$,
the spectra of $M(\nabla )M^{\dagger}(\nabla )$ and $M^{\dagger}(\nabla
)M(\nabla )$ are
identical, continuous and gapless. Outside the region, where $-\nabla > h$,
the spectra of $M(\nabla )M^{\dagger}(\nabla )$ and $M^{\dagger}(\nabla
)M(\nabla )$ are
identical, continuous and there is a gap. All unwanted ``doubler modes'' of the
na\"{\i}ve fermionic
action are in this third region. Note that the abrupt disappearance of the
zero mode at the boundary between
the two regions is a necessary consequence of the generic structure of the
supersymmetric
quantum mechanics governing the internal space and not of the specific choices
we have made.
In short, the trick in the lattice version amounts to walling a small
region of momentum space around the origin off the dangerous areas inhabited by
doublers.
The spectrum has a nonanalyticity
in momentum space as a result and one has to check carefully
for effects caused by this\footnote{*}{Of course, some singularity is needed
to avoid doublers in a non--dynamical way. }. Clearly, these nonanalyticities
come about because internal
space is infinite. The divergence responsible for the
singularity on the four dimensional Fourier torus is an infrared problem from
the point of view of the
extra dimension. Other approaches \cit{WEINSTEIN,REBBI}
that tried to use singularities
fail because gauge invariance forces compensating singular changes in the gauge
fermion
vertices \cit{KARSTENSMIT,PELISSETTO,BODWKO} and these introduce unwanted
doublers, ghosts, Lorentz violations or uncomputable non--perturbative effects
\cit{BODWKO,RABIN,WEINSTEINSCHW} into the theory.
Somehow, this new trick has to avoid all of these traps.

We know that the zero modes of $M$ live in the vicinity of the wall in the five
dimensional world and that the physics we are interested in is the defect
dependent part of
the action.
To eliminate five dimensional bulk effects the effective gauge interaction
induced by fermions, $S_{\rm eff} (U, wall)$, is
replaced by $S_{\rm eff} (U) =S_{\rm eff} (U,wall)
-{{S_{\rm eff}^+ (U) + S_{\rm eff}^- (U)}\over 2}$ where the $S_{\rm eff}^\pm
(U)$ come from systems
with constant mass terms equal to $\pm m_0$. The needed subtractions can also
be interpreted
as representing ghost fields. In analytical
computations, both in weak $\it and$ in strong coupling expansions, the needed
subtractions should be implementable order by order.

We make our remaining choices
so that the homogeneous systems which
provide the subtractions be as symmetric as
possible.
We would like to have  reflection positivity in the five dimensional system in
all
directions in the homogeneous case and five dimensional cubic symmetry when
$U=1$. Clearly,
the relevant five dimensional systems have the same Lagrangian as above, only
the matrix
$M^0$ gets replaced by $M^{\pm}$ with the corresponding $a^0 (s)$ replaced by
$a^{\pm} (s) =1 \pm  m_0 $. The five dimensional symmetries we want are then
obtained by making in each case
$M(\nabla_U ) = M^{\sigma} + {{\nabla_U}\over 2}$ with $\sigma=0,\pm$. It is
easy to see that with this choice the origin of four
dimensional momentum space is correctly walled off with
$0 < h=2m_0 < 2$. Note that we have more or less naturally ended up in the
$r=1$ case of the Wilson mass term (for notations see \cit{SMITACTA});
this case has better positivity properties than cases
where $|r|\ne 1$ because it has a positive bounded transfer matrix
for single step translations \cit{PELISSETTOMENOTTI,CREUTZ,LUSCHER}
so we should feel no need to generalize \cit{KAPLAN,JANSENSCHM,
KAPLANGOLT}.

The candidates for observables in the target theory are obtained as follows:
Pure gauge field operators are na\"{\i}vely transcribed. Local four
dimensional fermions are represented by a sum over $s$ of the bare fermion
fields with a weight that samples a finite but wide region around $s=0$.
The width of the weight functions should be taken to
infinity at the very end. Composite
operators of the target theory
are best constructed directly in the regularized version, for example, the
currents associated with global symmetries will
have a sum over all $s$ in their definition via
Noether's formula. With this definition the argument that correct
chiral anomalies are obtained \cit{KAPLAN,KAPLANGOLT,
CALLANHARVEY, NACULICH, COSTELUSCHER, JANSEN} applies.
The contributions of the subtraction terms $S_{\rm eff}^{\pm} (U)$
to anomalies cancel against each other because of the opposite signs
of the mass terms. One expects
currents associated with anomalous global symmetries to be affected by the
infinite sum over $s$ in a non--trivial way, enabling the charges
to flow out to infinity and hence be nonconserved in any finite
five dimensional slab
containing the defect. This shows that is is unclear whether exact
unitarity holds
if one looks only at such arbitrarily large, but still finite, slabs. A proper
construction will
define the whole theory in such a slab with boundary conditions is the $s$
direction (the other directions are irrelevant and can be kept infinite or
made finite by standard periodic/antiperiodic boundary conditions) chosen in
such a way that
anomalous charges will be able to flow out of the system. If the gauged group
is not anomalous
one would like then to see that a finite limit is obtained when the slab's
thickness is taken
to infinity. The choice of good boundary conditions needs care because in a
finite dimensional
internal space the zero modes will typically appear both in $M$ and in
$M^{\dagger}$; this might
be avoidable if the boundary conditions replace $M$ and $M^{\dagger}$ by
matrices that are
not each other's adjoint but this would give up the unitarity of the theory for
any finite width of the
slab.

The advantage of the present approach is that one can set up a meaningful test
of the
construction within perturbation theory and this test can
be carried out before the issue of boundary conditions is settled since the
subtractions
remove large $s$ divergences and one can work always with an infinite
internal space. As in the Pauli--Villars case we consider the fermion
contribution to
the vacuum polarization; this only tests the correctness of $S_{\rm eff} (U)$
but is
known to be a test failed by all previous proposals. From the imaginary part of
the
polarization tensor at small momenta one can read off the spectrum of fermions
as seen
by the {\it gauge bosons} and this is the relevant question. It is obvious that
na\"{\i}vely the method
works because the propagator has the right structure at nearly zero momenta by
construction.
The only way this argument can fail in perturbation theory
is that there are further singularities in the integrand
of the relevant lattice Feynman diagrams and they make contributions to the
continuum limit.
If we think about the propagator in a basis of eigenstates of
$M^{\dagger}(\nabla) M(\nabla )$
and $M(\nabla ) M^{\dagger}(\nabla)$ there is a source of trouble as discussed
before. We
aim to show now that if one represents the propagators in four dimensional
lattice
Fourier space and in $s$--space one can perform the needed subtractions easily
and
no singularities beyond the ``good'' ones appear. In particular, none are
induced by carrying
out the $s$--space traces.

The propagator is given by a formula similar to eq.~\(FREEPROP), now on the
lattice and in Euclidean space. With $ \bar p_\mu =\sin (p_\mu )$ and
$\hat p_\mu = 2\sin{{p_\mu}\over {2}}$ we have:
$$
\eqalign{&
(-i\sum_\mu \gamma_\mu \bar p_\mu +
M^{\dagger}(\hat p ) )P_R G_R (p)
 +(-i\sum_\mu \gamma_\mu \bar p_\mu + M (\hat p ) )P_L G_L (p)
  \cr
 &G_R (p) ={1\over {\sum_\mu \bar p_\mu^2 + M(\hat p )M^\dagger (\hat p ) }},~~
 G_L (p)={1\over {\sum_\mu \bar p_\mu^2 + M^{\dagger}(\hat p )M  (\hat p ) }}
\cr}\eqno(PROPLATT1)
$$
$$
G_R (p) M(p)=M(p) G_L (p)
$$
$G_R(p)$ and $G_L(p)$ are inverses of second order difference
operators and an explicit expression can be obtained by standard
techniques. Both are symmetric under the interchange of $s$ and
$s'$ and for $s\ge s'$ they are given by
$$
{}~~~~~~~[G_R]_{ss'}(p)=
\cases{[A_R(p)-B(p)]e^{-\alpha_+(p)[s+s'+2]}+B(p)e^{-\alpha_+(p)[s-s']}&
for $s'\ge -1$\cr A_R(p)e^{-\alpha_+(p)[s+1]+\alpha_-(p)[s'+1]}& for
$s\ge -1\ge s'$\cr
[A_R(p)-C(p)]e^{-\alpha_- (p)[s+s'+2]}+
C(p)e^{-\alpha_-(p)[s-s']}& for $s\le -1$\cr}
$$
$$\eqno(SPECIAL)$$
$$
[G_L]_{ss'}(p)=
\cases{[A_L(p)-B(p)]e^{-\alpha_+(p)[s+s']}+B(p)e^{-\alpha_+(p)[s-s']}&
for $s'\ge 0$\cr A_L(p)e^{-\alpha_+(p)s+\alpha_-(p)s'}& for
$s\ge 0\ge s'$\cr
[A_L(p)-C(p)]e^{-\alpha_-(p)[s+s']}+C(p)e^{-\alpha_-(p)[s-s']}& for $s\le
0$\cr}
$$
where
$$0\le \alpha_\pm(p)={\rm arccosh}\Bigl\{  {1\over 2} \big [
a_\pm(p)+{1\over
a_\pm(p)}+ {\bar p^2\over a_\pm(p)}\big ] \Bigr\},~~~a_\pm(p)=1\mp m_0+{1\over
2}\hat p^2$$
$$A_R(p)={1\over a_- (p) e^{\alpha_- (p)}-a_+ (p) e^{-\alpha_+ (p)}},~~
A_L(p)={1\over a_+ (p) e^{\alpha_+ (p)}-a_- (p) e^{-\alpha_- (p)}}\eqno(G)$$
$$
B(p)={1\over 2a_+ (p) \sinh \alpha_+ (p)},~~
C(p)={1\over 2a_- (p) \sinh \alpha_- (p) }$$
The propagator needed for the subtraction terms are obtained by
replacing all $a_\pm$, $\alpha_\pm$ by either $a_+$, $\alpha_+$
or $a_-$, $\alpha_-$. Under each of these substitutions
$A_R(p)$, $A_L(p)$,
$B(p)$ and $C(p)$ all become equal to each other.
It is easy to see that $\alpha_\pm(p)>0$
for all $p$. Subtractions make the sum over $s,s'$ converge
exponentially and uniformly in the loop momentum $k$ and external
momentum $p$. All ordinary manipulations that prove the
transversality of
$\Pi_{\mu\nu}(p)$ and its symmetry properties $\Pi_{\mu\nu}(p)=
\Pi_{\nu\mu}(p)=\Pi_{\mu\nu}(-p)$ hold after the subtractions.
All vertex factors are singularity free throughout the
Brillouin zone. The only singularities in a typical integrand then
come from the possible poles in the amplitudes
$A_R(p)$, $A_L(p)$, $B(p)$ and $C(p)$. Since we have arranged for
$M(\hat p)$ to have a zero mode, there is a singularity at $p=0$
in $A_L(p)$. It is easy to see from the expressions for the
amplitudes that this is the only singularity.
A simple calculation gives
$$A_L(p)={m_0 (4-m_0^2 )\over 4p^2}+\ {\rm regular\ terms}\eqno(POLE)$$
The residue of the pole is just the normalization
factor associated with the zero mode at $p=0$.
The usual continuum expressions should therefore emerge. We
end our perturbative analysis remarking
that all ingredients for the calculation of $\Pi_{\mu\nu}(p)$
are explicitly present and a meticulous calculation will make the above
argument foolproof.

We can also address the issue of how the approach presented here
avoids
the Nielsen-Ninomiya \cit{NIELSEN} theorem. To this end, consider integrating
all the fermion degrees of freedom except the righthanded fermion
at $s=0$. The kinetic energy term will be non-local and can be
extracted from the expression for the propagator at $s=s'=0$.
We find that the kinetic energy term is given by
$${1\over A_L(p)\bar p^2} i\sum_{\mu} [ \gamma_{\mu} \bar p_{\mu}] P_R
\eqno(KINETIC)$$
By eq.~\(POLE)  $A_L (p) \bar p^2$ goes to the normalization factor of
the zero mode and eq.~\(KINETIC) has
the expected zero at $p=0$. The factor
$\bar p^2$ in the denominator generates fifteen poles in
the Brillouin zone. Normally these poles would have
generated ghost contributions \cit{PELISSETTO} and
destroyed the theory.
But, in this
approach, integrating out the fermion degrees of freedom also
generates
pure gauge terms and these cancel the contributions from the ghosts.
In earlier attempts to solve the problem \cit{REBBI}, it was understood that
the
bgauge part of the action has to be suitably adjusted to cancel the
ghost contributions. The approach presented here can be thought of as
one way to systematically achieve the apparently difficult aim faced
before.

\head{Final Comments}

It should be clear from this paper that there is a large amount of latitude in
choosing $M$ and
the associate internal space. It is not clear at the moment whether sticking to
the ``wall''
situation is the best, but certainly it seems a worthwhile route to pursue, in
particular
because it gives a familiar picture for where the anomalous charge deficit is
going. The basic trick
however may be useful even if the ``wall'' realization turns out to fail
ultimately; in its essence
what the trick does is to provide the ``infinite hotel'' that has been argued
to be a necessity for the
existence of genuinely chiral fermions \cit{NIELSEN}.

\head{Acknowledgements} This research was supported in part by the DOE
under grant  \# DE-FG05-90ER40559.

\references

\refis{ROME} A. Borrelli, L. Maiani, G. C. Rossi, R. Sisto, M. Testa, \np B333,
1990, 335.

\refis{FSLAV} S. A. Frolov, A. A. Slavnov, SPhT/92-051, Saclay preprint, 1992.

\refis{KAPLAN} D. B. Kaplan, \pl B288, 1992, 342.

\refis{FADSLAV} L. D. Faddeev, A. A. Slavnov, ``Gauge Fields, an Introduction
to Quantum Field
Theory'', Second Edition, Addison--Wesley, (1991).

\refis{SVETLONG} A. C. Longhitano, B. Svetitsky, \pl B126, 1983, 259.

\refis{WITTEN}  E. Witten, \np B185, 1981, 513.

\refis{WEINSTEIN} S. D. Drell, M. Weinstein, S. Yankielowicz, \pr D14, 1976,
487.

\refis{REBBI} C. Rebbi, \pl B186, 1987, 200.

\refis{KARSTENSMIT} L. H. Karsten, J. Smit, \np B144, 1978, 536.

\refis{PELISSETTO} A. Pelissetto, \annp 182, 1988, 177.

\refis{BODWKO} G. T. Bodwin, E. V. Kovacs, \pr D35, 1987, 3198.

\refis{RABIN} J. M. Rabin, \pr D12, 1981, 3218.

\refis{WEINSTEINSCHW} M. Weinstein, \pr D26, 1982, 839.

\refis{SMITACTA} J. Smit, \acta, B17, 1986, 531.

\refis{PELISSETTOMENOTTI} P. Menotti, A. Pelissetto, \np B(Proc. Suppl.) 4,
1988, 644.

\refis{CREUTZ} M. Creutz, \pr D35, 1987, 1460.

\refis{LUSCHER} M. L\"{u}scher, \cmp 54, 1977, 283.

\refis{JANSENSCHM} K. Jansen, M. Schmaltz, UCSD/PTH 92-29, UCSD preprint, 1992.

\refis{KAPLANGOLT} M. F. L. Golterman, K. Jansen, D. B. Kaplan, UCSD/PTH 92-28,
UCSD
preprint, 1992.

\refis{CALLANHARVEY} C. G. Callan, Jr., J. A. Harvey, \np B250, 1985, 427.

\refis{NACULICH} S. G. Naculich, \np B296, 1988, 837.

\refis{COSTELUSCHER} A. Coste, M. L\"{u}scher, \np B323, 1989, 631.

\refis{JANSEN} K. Jansen, \pl B288, 1992, 348.

\refis{NIELSEN} H. B. Nielsen, M. Ninomiya, in ``Trieste Conference on
Topological Methods
in Quantum Field Theories'',  World Scientific, 1991.

\endreferences

\endit